# Is *p*-Type Doping in SeO$_2$ Feasible?

Zewen Xiao(肖泽文)[*]

Wuhan National Laboratory for Optoelectronics, Huazhong University of Science and Technology, Wuhan 430074, China

Email: zwxiao@hust.edu.cn

**ABSTRACT:**

The significance of *p*-type transparent oxide semiconductors (TOS) in the semiconductor industry is paramount, driving advancements in optoelectronic technologies for transparent electronic devices with unique properties. The recent discovery of *p*-type behavior in SeO$_2$ has stirred both interest and confusion in the scientific community. In this Letter, we employ density functional theory calculations to unveil the intrinsic insulating characteristics of SeO$_2$, highlighting substantial challenges in carrier doping. Our electronic structure analyses indicate that the Se 5s$^2$ states are energetically positioned too low to effectively interact with the O 2p orbitals, resulting in a valence band maximum (VBM) primarily dominated by the O 2p orbitals. The deep and localized nature of the VBM in SeO$_2$ limits its potential as a high-mobility *p*-type TOS. Defect calculations demonstrate that all intrinsic defects in SeO$_2$ exhibit deep transition levels within the bandgap. The Fermi level consistently resides in the mid-gap region, regardless of synthesis conditions. Furthermore, deep intrinsic acceptors and donors exhibit negative formation energies in the *n*-type and *p*-type regions, respectively, facilitating their spontaneous formation and impeding external doping efforts. Thus, the reported *p*-type conductivity in SeO$_2$ samples is unlikely intrinsic and is more plausibly attributed to reduced elemental Se, a well-known *p*-type semiconductor.

The research and development of *p*-type transparent semiconductors have become a critical focus in the semiconductor industry, driving advancements in optoelectronic technologies.[1–3] These materials are essential for transparent electronic devices, offering a unique combination of transparency and semiconductor functionalities crucial for various applications. However, the pursuit of high-mobility *p*-type transparent semiconductors, especially in the domain of oxides known as *p*-type oxide semiconductors (TOSs), faces significant challenges. Oxides are favored for their larger bandgaps ensuring transparency. Yet, a prevalent issue arises from the predominance of O 2p orbitals in the valence band maximum (VBM) of most oxides, hindering hole generation and transport.[4] To tackle this challenge, a common strategy involves utilizing cations with fully occupied $d^{10}$ or $s^2$ orbitals that can hybridize with O 2p orbitals, elevating and dispersing the VBM—a conventional approach in *p*-type TOS design.[3] The use of Cu(I) with high-lying Cu $3d^{10}$ orbitals has led to the development of Cu(I)-based *p*-type TOSs, exemplified by $CuAlO_2$.[3,5] Among common cations with $s^2$ configurations, the energy levels of $s^2$ orbitals deepen sequentially from Sn(II), Pb(II), Sb(III), Bi(III), Te(IV), to Po(IV).[6–8] SnO, characterized by a high-lying and dispersive VBM, stands out as a well-known *p*-type TOS.[9] In contrast, PbO presents a deeper and more localized VBM due to the relativistic effects of the Pb $6s^2$ orbitals, posing challenges in its classification as a *p*-type TOS.[10] Subsequently, the potential of $Sb_2O_3$, $Bi_2O_3$, $TeO_2$, and $PoO_2$ as *p*-type TOSs sequentially diminishes.

Interestingly, Zavabeti et al. have reported two-dimensional (2D) β-$TeO_2$ as a high-mobility *p*-type TOS,[11] sparking both interest and confusion.[12–16] However, recent work by Xiao et al.,[17] employing the doping limit rule,[18,19] decisively indicates that $TeO_2$, in bulk and 2D nanosheet forms,[20] is unlikely to be effectively doped *p*-type due to the too deep Te 5s orbitals. These orbitals exhibit minimal hybridization with the O 2p orbitals, failing to elevate the VBM and enhance *p*-type dopability. Xiao et al.[17] suggest that the observed *p*-type conductivity by Zavabeti et al. may arise from residual Se, potentially reduced Te, or $Te_{1-x}Se_x$ alloys, all recognized as high-mobility *p*-type semiconductors.[21–23]

Similarly, recent work by Qasrawi et al. has labeled $SeO_2$ as a *p*-type TOS,[24,25] generating interest and skepticism. It is noteworthy that due to Se being less oxophilic than Te,[11] $SeO_2$ is more susceptible to reduction to its elemental Se,[26] which also functions as a *p*-type semiconductor,[21] in contrast to TeO$_2$. The claimed *p*-type conductivity by Qasrawi et al. remains unverified, with the exact source of this behavior yet to be determined. To date, research on SeO$_2$ as a semiconductor is limited, and the dopability of SeO$_2$ has not been extensively examined.

In this Letter, we utilize density functional theory (DFT) calculations to demonstrate the intrinsic insulating behavior of SeO$_2$, highlighting significant challenges significant challenges for carriers doping. Electronic structure calculations reveal that the Se 5s$^2$ states in SeO$_2$ are situated even deeper than the Te 5s$^2$ states in TeO$_2$, thus they cannot effectively interact with the O 2p orbitals. As a result, the VBM is primarily composed of the O 2p orbitals. The deep and localized nature of the VBM in SeO$_2$ hinders its potential as a high-mobility *p*-type TOS. Defect calculations show that all intrinsic defects in SeO$_2$ exhibit deep transition levels within the bandgap. The Fermi level ($E_F$) consistently resides in the mid-gap region, irrespective of synthesis conditions. Additionally, deep intrinsic acceptors and deep intrinsic spontaneously form in the *n*-type and *p*-type regions respectively, rendering attempts at *n*-type or *p*-type external doping ineffective. Consequently, the observed *p*-type conductivity in SeO$_2$ samples could not be intrinsic and may originate from reduced elemental Se, a well-known *p*-type semiconductor.

DFT calculations were performed using the projector-augmented wave method implemented within the VASP 6.4.3 code.[27] A plane-wave cutoff energy of 500 eV and Γ-centered *k*-meshes with a *k*-point spacing of 0.2 Å$^{-1}$ were employed. All structures underwent full relaxation using the Perdew−Burke−Ernzerhof (PBE) functional[28] until the forces on each atom were below 0.01 eV/Å. Similar to TeO$_2$, SeO$_2$'s bandgap was underestimated by the PBE functional, with calculated and experimental values of 3.27 eV and 3.77 eV,[29,30] respectively. To address this, the Heyd−Scuseria−Ernzerhof (HSE) hybrid functional[31,32] with an optimized mixing parameter of 0.10 (HSE$^{0.10}$), providing a more accurate bandgap description (calculated value of 3.81 eV), was utilized for electronic structure and total energy calculations, similar to the case for TeO$_2$.[17] Defect calculations were conducted using a 2×2×1 supercell (containing 384 atoms) and a single Γ k-point. While maintaining the supercell lattice parameters constrained, atomic positions were relaxed using the PBE functional until the forces on each atom were below 0.03 eV/Å. The total energies of were calculated using the HSE$^{0.10}$ functional.

The formation energy $\Delta H_{\alpha,q}$ of a defect ($\alpha$) in a charge state $q$ was determined by the equation[33]

$$\Delta H_{\alpha,q} = E_{\alpha,q} - E_h + q(E_v + E_F) + \sum_i n_i \mu_i, \quad (1)$$

where $E_{\alpha,q}$ is the total energy of the supercell with the defect ($\alpha$) in the charge $q$, obtained by the self-consistent potential correction (SCPC) method[34] within the VASP 6.4.3; $E_h$ is the total energy of the perfect supercell; $E_F$ is the Fermi level referred to VBM, $E_V$; $n_i$ indicates the number of $i$ atom added ($n_i < 0$) or removed ($n_i > 0$) when a defect is formed, and $\mu_i$ is the chemical potential of the $i$ atom which can be expressed with respect to that of an element phase ($\mu_i^{el}$) by $\mu_i = \mu_i^{el} + \Delta \mu_i$. Utilizing the calculated $\Delta H_{\alpha,q}$ values, the equilibrium $E_F$ ($E_{F,e}$) was ascertained through the self-consistent solution of semiconductor statistic equations, adhering to the charge neutrality condition, as elaborated in the literature.[35]

Illustrated in Fig. 1, SeO$_2$ crystallizes within the tetragonal system (space group $P4_2/mbc$), characterized by lattice constants $a$ = 8.36 Å and $c$ = 5.06 Å.[36,37] The O atoms exhibit two distinct types both crystallographically and chemically. The first type of O atom is bonded to two Se atoms and denoted as bridging O1, while the O atoms of the second type is bonded to a single Se atom and labeled as terminal O2. The Se atom is bonded to two bridging O1 atoms (with a bond length of 1.79 Å) and one terminal O2 atom (with a bond length of 1.62 Å), forming a SeO$_3$ pyramid. Describing the Se environment as irregularly tetrahedral, it consists of three O atoms and one lone pair (E) at the vertices. These SeO$_3$E tetrahedra are interconnected through corner sharing, resulting in infinite isolated chains along the $c$ axis. The projections of the chain along the $c$ axis (Fig. 1a) and the [110] direction (Fig. 1b) reveal a W-like shape and a zigzag shape, respectively. The chains are bound together by the electrostatic attraction of the O atoms in one chain to the Se atoms in another chain, leading to a 4-fold arrangement for the W-shaped chains. Each chain is surrounded by four neighboring chains, where two chains have Se atoms attracted to the terminal O2 atom (side of the W-type) of that chain, while the other two chains have terminal O2 atoms attracted to the Se atom (bottom of the W-type) of that chain. To facilitate later discussion later, the former and the latter are respectively designated as "side" and "bottom" neighboring chains.

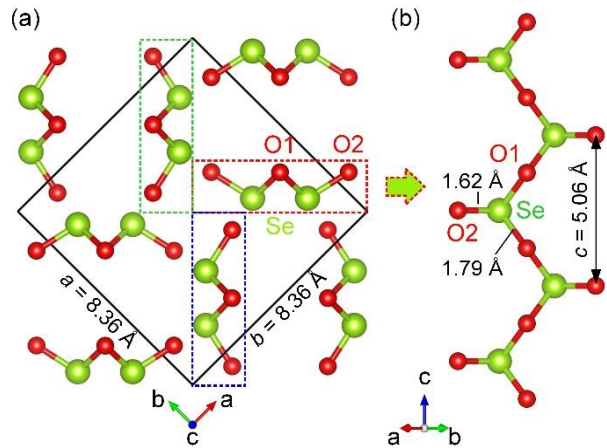

**Fig. 1.** Crystal structure of SeO$_2$ (space group $P4_2/mbc$): (a) the view along the $c$ axis and (b) the view along the [110] direction for the isolated W-shaped chain in the red dashed rectangle. For the chain within the red dashed rectangle, the chains within the green and blue dashed rectangles are respectively labeled as its "side" and "bottom" neighboring chains.

Before delving into the electronic structure of SeO$_2$, we first examine the chemical trends in the energies of the outermost s and p atomic orbitals of the p-block elements. In Fig. 2a, as we move across a period from left to right, the increasing effective nuclear charge causes a successive decrease in the energies of the outermost s and p atomic orbitals. Illustrated in Fig. 2b, descending down a group, with increasing atomic radius, the energies of the outermost p orbitals rise successively, while the energies of the outermost s orbitals exhibit a general upward trend with irregularities. For example, due to the influence of the 3d orbitals, the energy of the 4s orbital of Period 4 elements deviates from the expected trend (e.g., Ga 4s orbital is lower than Al 3s orbital). Additionally, as discussed by Walsh et al. in their review,[6] relativistic effects cause the 6s orbitals of Period 6 elements to be lower than the 5s orbitals of corresponding Period 5 elements. Thus, the pattern emerges as previously described, with the energy levels of the outermost s orbitals decreasing sequentially from In, Tl, Sn, Pb, Sb, Bi, Te to Po.

The energy relationship between the outermost s orbitals of p-block elements and the O 2p orbital significantly influences the electronic structures of their oxides. A high energy level of the outermost s orbital leads to electron loss, contributing to the conduction band, as observed in *n*-type In$_2$O$_3$. Conversely, slightly lower energies allow the s orbitals to remain filled and hybridize with the O 2p orbitals, contributing to the high-lying and delocalized VBM, as seen in *p*-type SnO.[9] However, when the energy of the outermost s orbital is excessively low, effective hybridization with the O 2p orbitals to achieve *p*-type dopability is hindered, as evidenced in insulating PbO[10] and TeO$_2$.[17,38,39] From Fig. 2, it is evident that the Se 4s orbital is lower in energy than the Te 5s orbital, suggesting that the Se 4s orbital is unlikely to contribute to the VBM, thus not promoting *p*-type dopability.

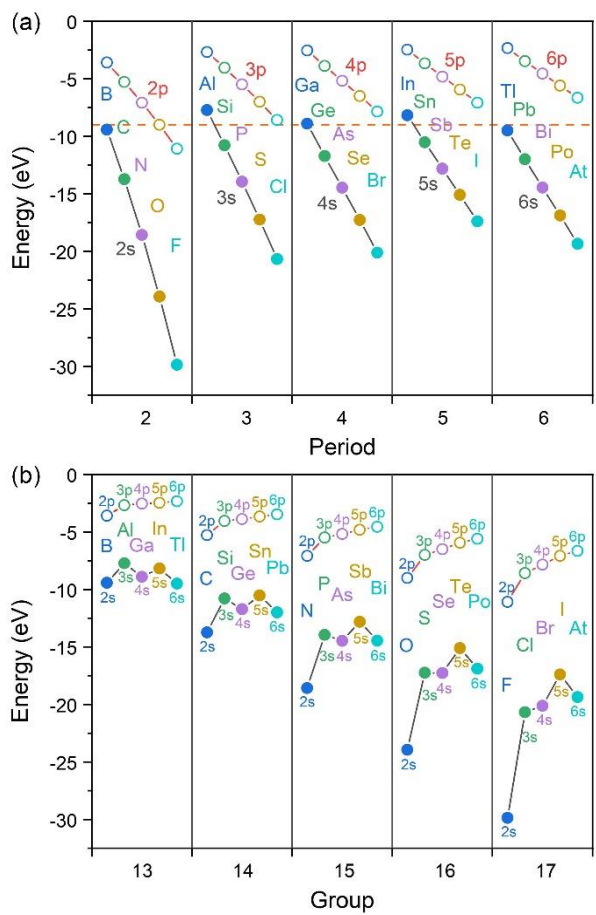

**Fig. 2.** Calculated energies (with respect to the vacuum level) of the outermost s and p atomic orbitals of p-block elements except for noble gases, organized by (a) period and (b) group. Note that the Se 4s orbital is deeper than the Te 5s orbital, thus further away from the O 2p orbital; the Se 4p orbital is deeper than the Te 5p orbital, thus closer to the O 2p orbital.

We now delve into the electronic structure of SeO$_2$. Illustrated in Fig. 3a, the band structure of SeO$_2$ unveils an indirect bandgap with the VBM located at the Z point and the conduction band minimum (CBM) at the M point. As expected, the Se 4s states are positioned at energy levels too low to interact with the O 2p orbitals, resulting in the VBM primarily composed of the O 2p orbitals. Fig. 3c illustrates that the charge density associated with the VBM is predominantly concentrated around the O atoms. Consequently, the VBM is relatively low-lying and localized, leading to large effective hole masses, measuring $2.16m_0$ ($m_0$ is the free electron mass), $10.22m_0$, and $9.42m_0$ along the Z–Γ (parallel to the chain direction), Z–R, and Z–A directions, respectively. These characteristics impede hole generation and transport in $SeO_2$. On the contrary, due to the proximity of the Se 4p orbitals to the O 2p orbitals (Fig. 2), they can interact effectively. The CBM comprises the antibonding states of the Se 4p and O 2p orbitals, as evidenced by the charge density associated with the CBM in Fig. 3b. The bonding states of the Se 4p and O 2p orbitals contribute to the lower segment of the valence band, highlighting the robust covalent nature of the Se—O bonds. The calculated effective electron mass is lower than the effective hole mass, measuring $0.22m_0$, $1.58m_0$, and $1.62m_0$ along the M–A (parallel to the chain direction), M–X, and M–Γ directions, respectively. In summary, the electronic structure of $SeO_2$ does not favor its potential as a high-mobility $p$-type TOS.

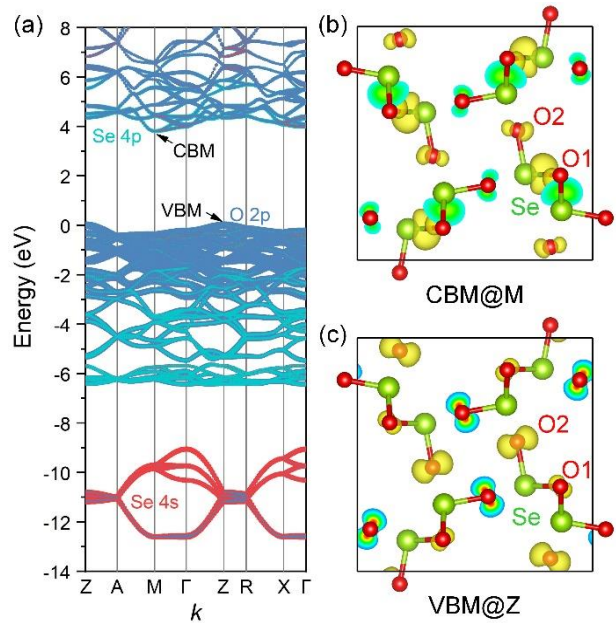

**Fig. 3.** Electronic structure of $SeO_2$: (a) band structure and (b,c) charge densities viewed along the $c$ axis for (b) the CBM at the M point and (c) the VBM at the Z point. Note that the low-lying Se 4s orbital does not participate in hybridization with the O 2p orbitals, resulting in no contribution to the VBM.

To assess the dopability of $SeO_2$, we investigated the thermodynamics of its intrinsic defects. Figure 4 illustrates the charge transition levels of all intrinsic defects. For the predominant vacancy defect, we compared its defective local structures with the perfect structure (Fig. 5a), as shown in Fig. 5. When a neutral Se vacancy ($V_{Se}$) is generated, three Se—

O bonds break. Subsequently, the terminal O2 atom relocates to the vacancy site and bonds with two bridging O1 atoms (O2—O1 bond length: 1.44 Å, slightly longer than the calculated O—O bond lengths in $O_2$ and $O_3$ molecules at 1.23 and 1.28 Å, respectively), as depicted in Fig. 5b. Upon gaining two electrons, the three O atoms surrounding $V_{Se}$ disperse, with the terminal O2 atom bonding to the Se atom on the side neighboring chain (bond length: 1.62 Å), as shown in Fig. 5c. The (0/2–) transition of $V_{Se}$ occurs significantly within the mid-gap, specifically at 1.75 eV above the VBM (Fig. 4), indicating that $V_{Se}$ is not an effective acceptor. The deep nature of the (0/2–) transition of $V_{Se}$ is consistent with the negligible contribution of Se 4s to the VBM, reflecting the strong covalency between Se and O. With the $E_F$ positioned above the VBM by 2.11 eV, $V_{Se}$ acquires two additional electrons, resulting further dispersion of the three O atoms around it. The terminal $O_2$ atom disrupts the side neighboring chain and forms bonds with two Se atoms in that chain (bond length: 1.91 Å), as shown in Fig. 5d. Conversely, as the $E_F$ drops below 0.63 eV above the VBM, $V_{Se}$ loses two electrons, causing the three surrounding O atoms to cluster, forming an O1—O2—O1 trimer akin to an $O_3$ molecule, as depicted in Fig. 5e. Therefore, in the p-type region, $V_{Se}$ transforms into an acceptor killer.

There are two types of O vacancies. In the case of the bridging O1 vacancy ($V_{O1}$), when neutral, the two adjacent unbridged Se atoms reconnect by forming a Se—Se bond (bond length: 2.78 Å), as shown in Fig. 5f. Upon losing two electrons, the positively charged Se atom moves away and bond with the O2 atom on the bottom neighboring chain (bond length: 1.94 Å), as depicted in Fig. 5g. The (0/+2) transition of $V_{O1}$ lies significantly below 2.30 eV from the CBM, indicating its inefficiency as a donor. Even in the n-type region, $V_{O1}$ acquires two electrons, and the terminal O2 atoms, connected to two Se atoms, bond with Se atoms on the side neighboring chains (bond length: 2.06 Å). Likewise, the terminal O2 vacancy ($V_{O2}$) proves ineffective as a donor, with the (0/2+) transition positioned at 2.41 eV below the CBM. In the case of neutral $V_{O2}$, the 2-coordianted Se atom moves slightly towards the two bridging O1 atoms, resulting in a slightly reduced bond length of 1.82 Å (Fig. 5i). For doubly positively charged $V_{O2}$, the Se atom beside the vacancy forms an additional bond with the O2 atom on the bottom neighboring chain (bond length: 1.87 Å) (Fig. 5j). As summarized in Fig. 4, the transition levels of other intrinsic defects also fall within the mid-gap, rendering them ineffective as donors or acceptors.

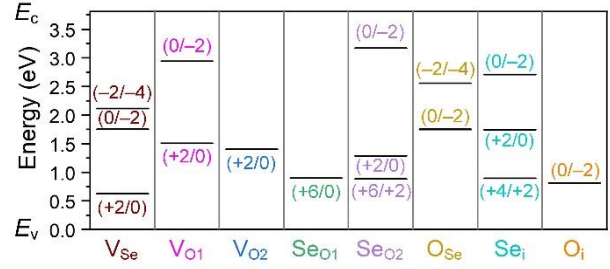

**Fig. 4.** Calculated charge transition levels of intrinsic defect in $SeO_2$. "$E_v$" and "$E_c$" represent the VBM and CBM, respectively. Note that all intrinsic defects in $SeO_2$ exhibit deep transition levels.

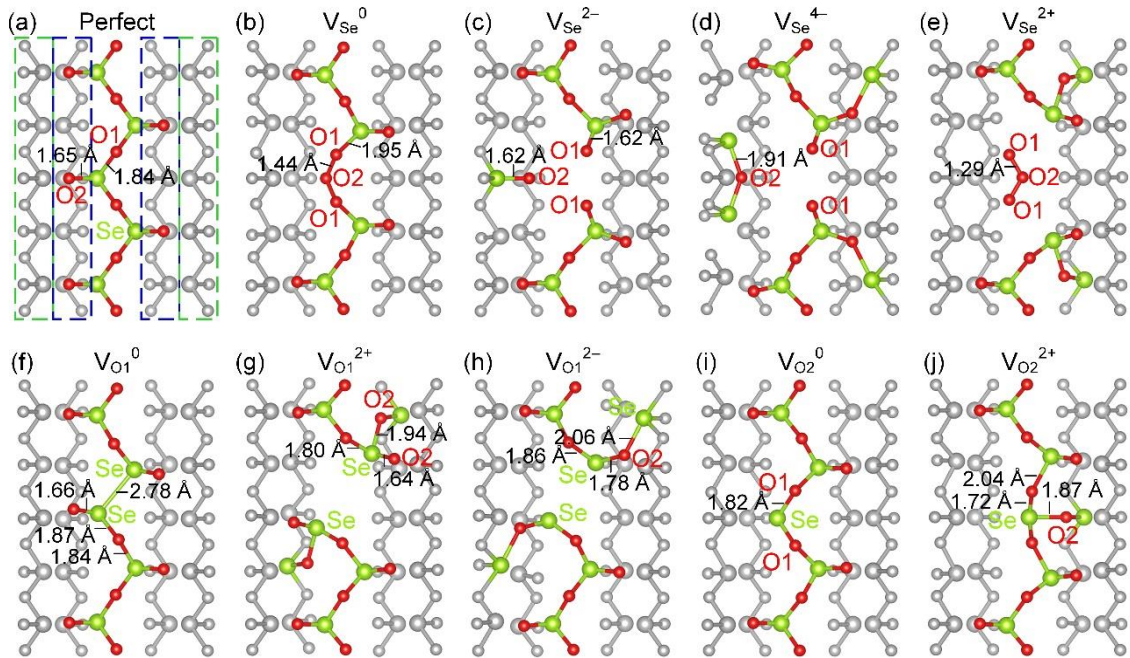

**Fig. 5.** Relaxed local structures of (a) perfect structure, (b) $V_{Se}^0$, (c) $V_{Se}^{2-}$, (d) $V_{Se}^{4-}$, (d) $V_{Se}^{2+}$, (e) $V_{Se}^0$, (f) $V_{O1}^0$, (g) $V_{O1}^{2+}$, (h) $V_{O1}^{2-}$, (i) $V_{O2}^0$, and (j) $V_{O2}^{2+}$. The atoms within the green and blue dashed boxes represent the atoms in the "side" and "bottom" neighboring chains, respectively. For clarity, these atoms are depicted in grey unless they are bonded to the atoms on the defective chain.

Figure 6 presents the calculated formation energies ($\Delta H$) of intrinsic defects in $SeO_2$ as a function of the $E_F$. In the Se-rich/O-poor condition, delineated by the phase equilibrium between $SeO_2$ and elemental Se, $V_{O1}$ and $V_{O2}$ showcase the lowest $\Delta H$ values. Additionally, the two Se-on-O antisites ($Se_{O1}$ and $Se_{O2}$) exhibit relatively low $\Delta H$ values. However, these four defects stabilize in neutral states. Consequently, the calculated equilibrium $E_F$ ($E_{F,e}$) at 300 K positions 2.09 eV above the VBM (i.e., 1.72 eV below the CBM), corresponding to zero carrier concentration, signifying insulating characteristics. Despite elevating in the growth temperature ($T_G$) of $SeO_2$ to 600 K (melting point at 613 K), the $E_{F,e}$ remains relatively constant, indicating the sustained insulating nature of $SeO_2$. Under the Se-poor/O-rich condition, governed by the phase equilibrium between $SeO_2$ and $O_2$ (considering the metastable phases $Se_2O_5$[40] and $SeO_3$[41] above the hull), $V_{Se}$ and O-on-Se antisites ($O_{Se}$) exhibit the lowest $\Delta H$ values, stabilizing in neutral states. Despite the calculated concentrations of $V_{Se}$ and $O_{Se}$ reaching up to $4.9 \times 10^{18}$ cm$^{-3}$ and $9.4 \times 10^{19}$ cm$^{-3}$ respectively, the calculated $E_{F,e}$ stands at 1.18 eV above the VBM, with negligible hole concentrations (<$10^1$ cm$^{-3}$) and zero electron concentration. Modulating the $T_G$ minimally impacts the $E_{F,e}$ and the insulating properties. These results suggest that regardless of growth chemical conditions and temperature variations, $SeO_2$ will consistently demonstrate insulating traits.

Notably, in both Se-rich/O-poor and Se-poor/O-rich scenarios, within the *n*-type region, the $\Delta H$ values of negatively charged deep acceptors are negative, while in the *p*-type region, the $\Delta H$ values of positively charged deep donors are negative. This indicates the spontaneous formation of compensating defects, impeding $E_{F,e}$ from entering the *n*-type or *p*-type regions, rendering attempts at *n*-type or *p*-type external doping ineffective.

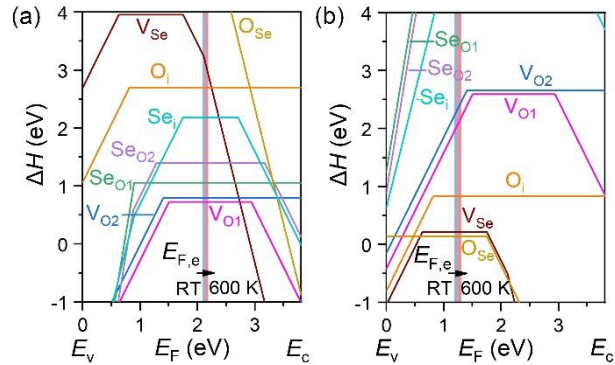

**Fig. 6.** Calculated formation energies ($\Delta H$) of intrinsic defects in $SeO_2$ as a function of the Fermi level ($E_F$) under (a) the Se-rich/O-poor condition (i.e., $\Delta\mu_{Se} = 0$ eV and $\Delta\mu_O = -1.87$ eV) and (b) the O-rich/Se-poor condition (i.e., $\Delta\mu_{Se} = -3.74$ eV and $\Delta\mu_O = 0$ eV). The colored bars represent the ranges of equilibrium $E_F$ ($E_{F,e}$) solved with growth temperatures ($T_G$) ranging from room temperature to a sufficient high temperature of 600 K (melting point at 613 K). The $E_{F,e}$ is situated at the mid-gap region, irrespective of the chemical potential and $T_G$, indicating constant insulating properties.

In summary, our DFT investigation of the electronic structure and dopability of $SeO_2$ highlights its intrinsic insulating nature and the substantial hurdles associated with carrier doping. Our electronic structure analyses reveal that the Se $5s^2$ states reside at energy levels too low to effectively hybridize with the O 2p orbitals, resulting in a VBM primarily composed of the O 2p orbitals. The deep and localized VBM of $SeO_2$ diminishes its potential as a high-mobility *p*-type TOS. Defect calculations demonstrate that all intrinsic defects in $SeO_2$ feature deep transition levels in the bandgap. The $E_{F,e}$ remains consistently within the mid-gap region, regardless of synthesis conditions. Additionally, in the *n*-type and *p*-type regions, deep intrinsic acceptors and deep intrinsic donors possess negative $\Delta H$ values, respectively, leading to their spontaneous formation, thereby rendering attempts at *n*-type or *p*-type external doping ineffective. Consequently, the observed *p*-type conductivity in $SeO_2$ samples is unlikely to be intrinsic and is more likely linked to the reduced elemental Se, a well-established *p*-type semiconductor.

*Acknowledgements*. This work was financially supported by the National Natural Science Foundation of China (grant no. 52372150).